\documentclass[useAMS,usenatbib]{mn2e}

\usepackage{graphicx}
\usepackage{color}
\usepackage{amsmath, amssymb}
\usepackage{natbib}

\usepackage{ulem}

\title[Evolution of 3-dimensional Relativistic Current Sheets]
{Evolution of 3-dimensional Relativistic Current Sheets 
 and Development of Self-Generated Turbulence}

\author[Takamoto]{M. Takamoto,$^{1}$\thanks{E-mail:
mtakamoto@eps.s.u-tokyo.ac.jp}
\\
$^{1}$Department of Earth and Planetary Science, University of Tokyo, Tokyo 113-0033, Japan \\
}
\begin{document}

\date{}

\pagerange{\pageref{firstpage}--\pageref{lastpage}} \pubyear{2017}

\maketitle

\label{firstpage}

\begin{abstract}
In this paper, 
the temporal evolution of 3-dimensional relativistic current sheets in Poynting-dominated plasma is studied 
for the first time. 
Over the past few decades, 
a lot of efforts have been conducted on studying the evolution of current sheets in 2-dimensional space, 
and concluded that sufficiently long current sheets always evolves into the so-called ``plasmoid-chain'', 
which provides fast reconnection rate independent of its resistivity. 
However, 
it is suspected that plasmoid-chain can exist only in the case of 2-dimensional approximation, 
and would show transition to turbulence in 3-dimensional space. 

We performed 3-dimensional numerical simulation of relativistic current sheet using resistive relativistic magnetohydrodynamic approximation. 
The results showed that 
the 3-dimensional current sheet evolve not into plasmoid-chain but turbulence. 
The resulting reconnection rate is $0.004$ which is much smaller than that of plasmoid-chain. 
The energy conversion from magnetic field to kinetic energy of turbulence is just 0.01\% 
which is much smaller than typical non-relativistic cases. 
Using the energy principle, 
we also showed that 
the plasmoid is always unstable for a displacement in opposite direction to its acceleration, probably interchange-type instability, 
and this always results in seeds of turbulence behind the plasmoids. 
Finally, 
the temperature distribution along the sheet is discussed, 
and it is found that the sheet is less active than plasmoid-chain. 
Our finding can be applied for many high energy astrophysical phenomena, 
and can provide a basic model of the general current sheet in Poynting-dominated plasma. 
\end{abstract}


\section{Introduction}
\label{sec:sec1}

Recently, 
the development of many high energy astronomical observation devices has allowed us to find many flare phenomena from various high energy astrophysical phenomena, 
such as Crab pulsar wind \citep{2012ApJ...749...26B,2013MNRAS.436L..20B,2014RPPh...77f6901B,2015PPCF...57a4034P,2015MNRAS.454.2972T} 
and blazars \citep{2012ApJ...754..114H,2015ApJ...808L..18A}. 
Relativistic magnetic reconnection is considered to be a good candidates for those phenomena. 
This is because magnetic reconnection efficiently converts the magnetic field energy into plasma kinetic, thermal, photon, and non-thermal particle energy. 
In addition, 
it is known that 
sufficiently long current sheets in 2-dimensional space always evolves into the so-called ``plasmoid-chain'' 
in which the current sheets are filled with many plasmoids generated by the secondary tearing instability \citep{2001EP&S...53..473S,2005PhRvL..95w5003L,2007PhPl...14j0703L}. 
The plasmoids experience many collisions to the neighboring ones, 
and it is expected that 
the energy released by such collisions can be responsible for flare phenomena observed in the high energy astrophysical phenomena. 

The research of magnetic reconnection in relativistic plasma, in particular, in Poynting-dominated plasma 
has been conducted vigorously for these decades. 
After several initial analytic work \citep{1994PhRvL..72..494B,2003MNRAS.346..540L,2003ApJ...589..893L,2005MNRAS.358..113L}, 
numerical simulation became the main method for studying relativistic magnetic reconnection due to its strong non-linear effects. 
Using relativistic magnetohydrodynamic (RMHD) approximation, 
\citet{2006ApJ...647L.123W,2011ApJ...739L..53T} have studied the initial phase of the tearing instability in current sheets with low Lundquist number, 
$S \equiv 4 \pi L_{\rm sheet} c_{\rm A}/\eta \lesssim 10^3$ 
where $L_{\rm sheet}$ is the sheet length, $c_{\rm A}$ is the Alfv\'en velocity, and $\eta$ is the resistivity. 
They observed strong compression in the downstream region predicted by \citet{1994PhRvL..72..494B,2003ApJ...589..893L,2005MNRAS.358..113L}, 
though the observed reconnection rate is very similar to the non-relativistic case predicted by \citet{2003ApJ...589..893L,2005MNRAS.358..113L}. 
However, 
it has been shown that 
relativistic magnetic reconnection results in faster reconnection rate than the non-relativistic case 
in Poynting-energy dominated plasma 
if much larger Lundquist number, $S > 10^4$, is considered and the sheet evolved into ``plasmoid-chain'' \citep{2013ApJ...775...50T}. 
On the other hand, 
there are also several work taking into account plasma effects, such as considering two-fluid approximation and fully collisionless plasma. 
\citep{2009ApJ...696.1385Z,2009ApJ...705..907Z} have performed numerical simulations of relativistic magnetic reconnection 
by assuming relativistic two-fluid approximation, 
and observed enhancement of reconnection rate as electromagnetic energy increases.
A similar effect were later observed in resistive RMHD simulation of Petschek reconnection \citep{2010ApJ...716L.214Z}. 
Collisionless reconnection work using Particle-in-cell (PIC) have also performed in this decade \citep{2014ApJ...783L..21S,2015PhRvL.114i5002L,2015ApJ...806..167G}. 
In addition to the increase of reconnection rate as electromagnetic field energy, 
they found that relativistic magnetic reconnection in Poynting-energy dominated plasma is a very efficient accelerator of particles, 
and can be a good candidate for flare events of high energy astrophysical phenomena. 

In spite of the above very active researches, 
there are still only a few work of magnetic reconnection in 3-dimensional space. 
It is considered that 
the current sheet in 3-dimensional scale will not evolve into plasmoid-chain but turbulence 
because of many instabilities in current sheet 
which can break the symmetry assumed in 2-dimensional work. 
In this case, 
it is known that turbulence enhances the magnetic reconnection rate. 
One reason of this is that turbulent motion of magnetic field increases the dissipation around reconnection point \citep{2013PhRvL.110y5001H}; 
In addition, more importantly, 
it was shown that 
turbulent eddy motion drives diffusion of magnetic field line separation, 
resulting in broader exhaust region and faster reconnection rate \citep{1999ApJ...517..700L,2009ApJ...700...63K,2015ApJ...815...16T}
The above work assumed an external driven turbulence, 
and it depends on each phenomenon if there is sufficiently strong turbulence in those environments. 
However, 
current sheets have various kinds of instability, 
and it is expected that such instabilities will evolve into turbulence. 
Hence, 
the recent main research interest is to find detailed mechanisms of the self-generated turbulence in sheet, 
and the resulting reconnection rate. 
In non-relativistic case, 
there are a few work on self-generating turbulence in current sheets \citep{2015ApJ...806L..12O,2016ApJ...818...20H,2017ApJ...838...91K}. 
They reported that the 3-dimensional self-generate turbulent current sheets show smaller reconnection rate $\sim 0.005 - 0.01$, 
and 1 to 5 \% of the magnetic field energy conversion into kinetic energy of the turbulence. 
%

In this paper, 
we report the first study of the temporal evolution of relativistic 3-dimensional current sheets in Poynting-dominated plasma 
using a new Godunov type scheme of resistive relativistic magnetohydrodynamic simulation. 
Since recent studies showed that 
turbulence in Poynting-energy dominated plasma has very different properties from non-relativistic one \citep{2015ApJ...815...16T,2016ApJ...831L..11T,2017MNRAS.472.4542T}, 
we expect that such effects may modify the behavior of self-generated turbulence in relativistic sheets. 
In Section~2 we introduce the numerical setup. 
The numerical result is presented in Section~3, 
and its theoretical discussion is given in Section~4. 
Section~5 summarizes our conclusions. 

\section{Numerical Setup}
\label{sec:sec2}

In this paper, 
an evolution of a very long current sheet is modeled using the relativistic resistive magnetohydrodynamic approximation. 
We use a newly developed resistive relativistic magnetohydrodynamics (RRMHD) scheme explained in the Appendix, 
which is an extension of our previous work \citep{2011ApJ...735..113T}, 
and allows us to obtain the full-Godunov solver of RRMHD for the first time. 
We calculate the RRMHD equations in a conservative fashion, 
and the mass density, momentum, and energy are conserved within machine round-off error. 
We use the constrained transport algorithm \citep{1988ApJ...332..659E} 
to preserve the divergence free constraint on the magnetic field. 
The multi-dimensional extension is achieved using the unsplit method \citep{2005JCoPh.205..509G,2008JCoPh.227.4123G}. 
For the equation of state, 
a relativistic ideal gas with $h = 1 + (\Gamma / (\Gamma - 1))(p / \rho)$, $\Gamma = 4 / 3$ is assumed 
where $\rho$ is the rest mass density, and $p$ is the gas pressure in the plasma rest frame. 
The resistivity $\eta$ is determined from the Lundquist number: $S \equiv 4 \pi L_{\rm sheet} c_{\rm A}/ \eta = 2.912 \times 10^5$. 

For our numerical calculations, 
we prepare a numerical box, $[-L_{\rm x}, L_{\rm x}] \times [0, L_{\rm y}] \times [-L_{\rm z}, L_{\rm z}] = [-80 L, 80 L] \times [0, 20L] \times [-20L, 20L]$, 
where $L$ is the initial current sheet width. 
We divide it into homogeneous numerical meshes as $N_{\rm x} \times N_{\rm y} \times N_{\rm z} = 4096 \times 1024 \times 2048$. 

Along the boundaries at $x = \pm L_{\rm x}$ and $z = \pm L_{\rm z}$, 
the free boundary condition is imposed; 
Along the boundaries at $y = 0, L_{\rm y}$, 
the periodic boundary condition is imposed. 
For the initial condition, 
the static relativistic Harris current sheet \citep{1966PhFl....9..277H,2003ApJ...591..366K} is assumed as: 
\begin{equation}
  B_z(x)  =  B_0 \tanh (z / L)
  .
\end{equation}
The uniform temperature, $k_{\rm B} T = m c^2$, is assumed 
where $k_{\rm B}$ is the Boltzmann constant, 
$m$ is the particle rest mass, 
and $c$ is the light velocity. 
The gas pressure  satisfies the pressure balance condition, 
and the upstream gas pressure is determined by the magnetization parameter: $\sigma \equiv B_0^2 / 4 \pi \rho h c^2 \gamma^2 = 5$ 
where $\gamma$ is the Lorentz factor which is unity initially. 

To save numerical simulation time in 3-dimensional space, 
we first perform a 2-dimensional simulation. 
In this simulation, 
to trigger the initial tearing instability at the origin $(x,z) = (0,0)$, 
we add the following small perturbation to the magnetic field: 
\begin{equation}
  \delta A_{\rm y} = - 0.01 B_0 L \exp[-(x^2 + z^2) / 4 L^2]
  .
\end{equation}
The 2-dimensional simulation is performed until $t = 60 L/c$ 
when the initial tearing mode is fully developed 
but the secondary tearing instability is not developed. 
We copied the numerical result of this 2-dimensional run into y-direction, 
and added a small white noise perturbation around $(x, z) = (0, 0)$ in velocity and motional electric field: ${\bf E} = - {\bf v} \times {\bf B}$ 
whose velocity dispersion is 0.1c. 
Note that 
the above time, $t = 60 L/c$, is redefined to the initial time $t = 0$ of the 3-dimensional simulation and analysis in the following. 

In the following, 
we use the unit: $c = 1$. 

\section{Results}
\label{sec:sec3}

\begin{figure}
 \centering
  \includegraphics[width=8.cm,clip]{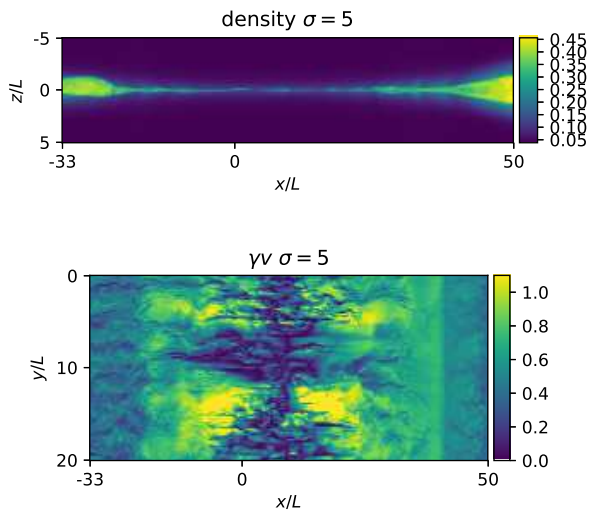}
  \caption{The profiles of rest mass density and $|\gamma v|$ at $t = 120 L/c$. 
           }
  \label{fig:0}
\end{figure}

Figure \ref{fig:0} is side view of density and top view of absolute value of velocity $|\gamma v|$ at $t = 120 L/c$. 
The top panel shows that 
there are fragmentation and small scale fluctuations. 
However, no clear small scale plasmoids are observed in the current sheet. 
In the bottom panel, 
it shows that 
there is no coherent structure in y-direction along the current sheet, 
indicating the appearance of fully evolved turbulence in the sheet. 
Interestingly, 
in the bottom panel 
we found that 
a region with relativistic velocity, close to the Alfv\'en velocity in the upstream region, is appeared locally in the sheet. 
this is in contrast to the results of 2-dimensional tearing instability in Poynting dominated plasma 
which resulted in a much slower outflow velocity, typically $\sim 0.3 c$. 

\begin{figure}
 \centering
  \includegraphics[width=8.cm,clip]{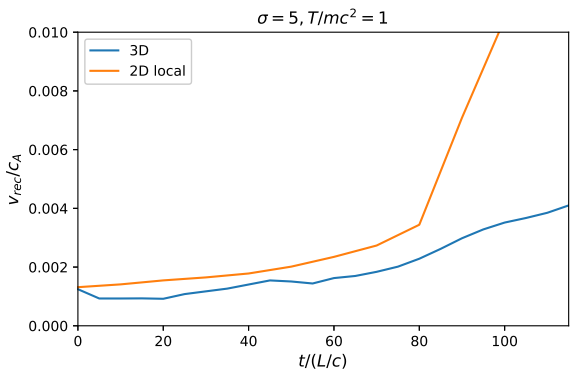}
  \caption{Temporal evolution of reconnection rate 
           in 2 and 3 dimensional cases. 
           }
  \label{fig:1}
\end{figure}

Figure \ref{fig:1} is the temporal evolution of reconnection rate obtained by our simulation in 2 and 3 dimensional cases. 
In 2-dimensional case, reconnection rate is measured as: 
\begin{equation}
  \label{eq:3.1}
  v_{\rm R} / c_{\rm A} \equiv \frac{1}{2 B_0 c_{\rm A} L_{\rm x}} \left | \int^{L_{\rm x}}_{-L_{\rm x}} dx E_{\rm y}(x, z=0)  \right|
  .
\end{equation}
In 3-dimensional case, 
reconnection rate is measured by a method proposed by \citep{2009ApJ...700...63K} 
which is a natural extension of Equation (\ref{eq:3.1}) into 3-dimensional space. 

Figure \ref{fig:1} shows that 
any clear differences cannot be observed until $t = 80 L/c$, 
and it shows linear and quasi-linear evolution of tearing instability in both 2 and 3 dimensional cases. 
At $t = 80 L/c$, 
reconnection rate of 2-dimensional case shows a rapid growth due to the evolution of plasmoid-chain, 
and it grows up to $0.03$ as observed in \cite{2013ApJ...775...50T}. 
The 3-dimensional case does not show such a rapid growth. 
Reconnection rate increases but more moderately 
whose saturation value seems around 0.004. 
Note that this value is similar to non-relativistic work \citep{2015ApJ...806L..12O,2016ApJ...818...20H,2017ApJ...838...91K}. 
We will discuss the meaning of this value later. 

\begin{figure}
 \centering
  \includegraphics[width=8.cm,clip]{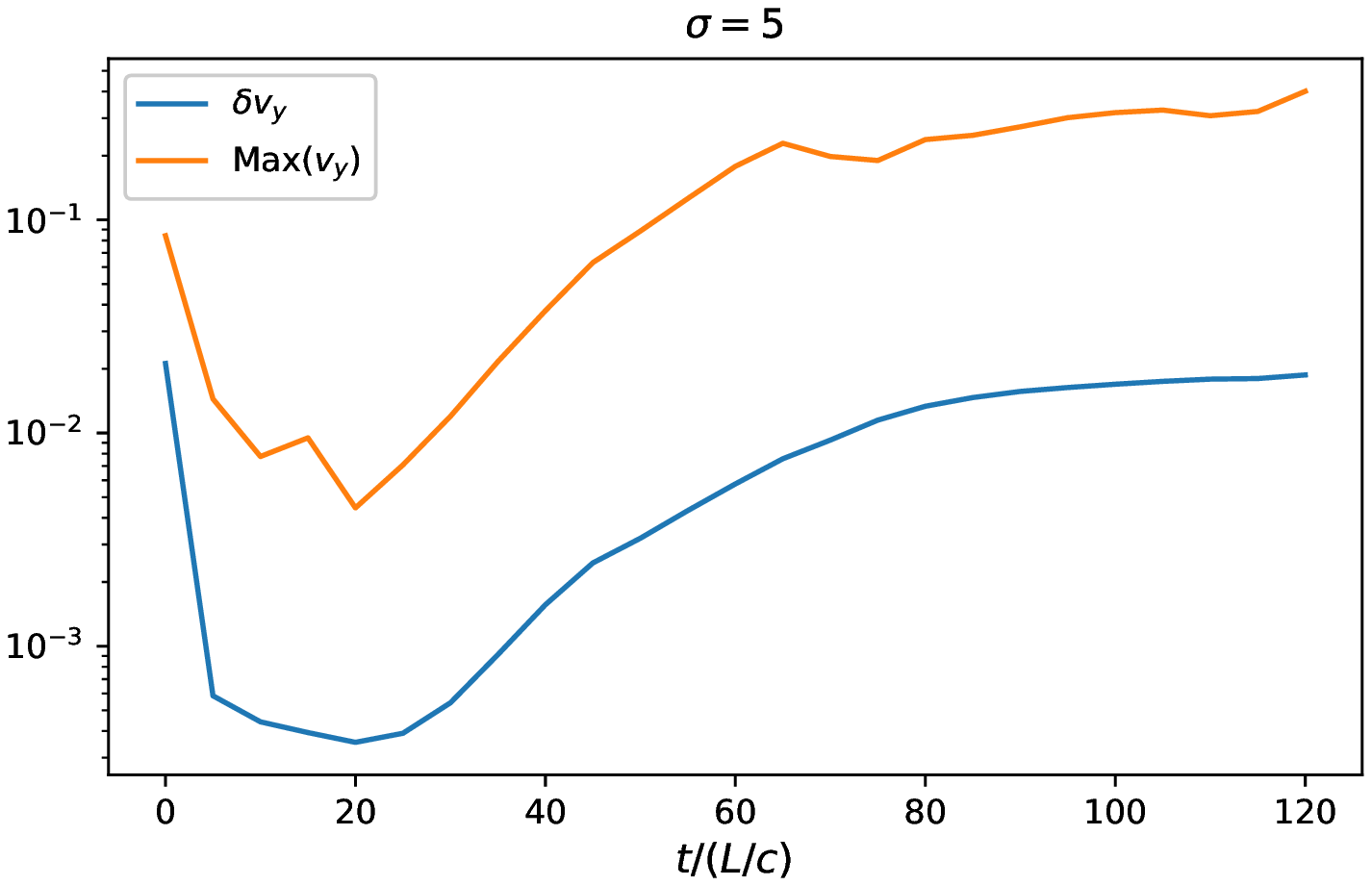}
  \includegraphics[width=8.cm,clip]{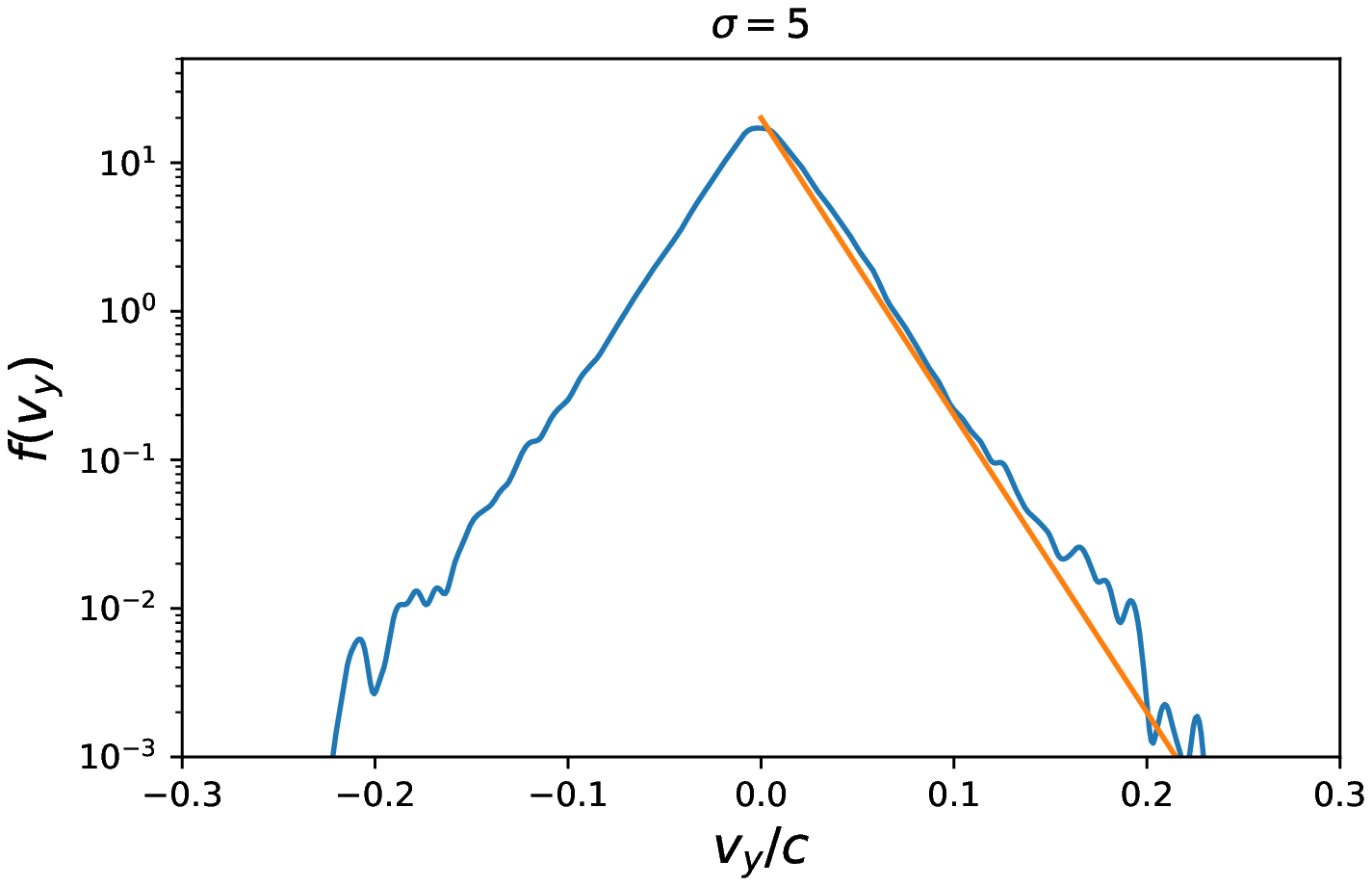}
  \caption{Top: Temporal evolution of maximum velocity and velocity dispersion of $v_{\rm y}$. 
           Bottom: Distribution function of $v_{\rm y}$ at $t = 120 L/c$. 
           The orange line is the fitting curve with the form: $\exp[- |v_{\rm y}|/\delta v_{\rm y}]$ 
           where $\delta v_{\rm y}$ is the velocity dispersion of $v_{\rm y}$. 
           }
  \label{fig:2}
\end{figure}

Top panel of Figure \ref{fig:2} is the evolution of maximum value and dispersion of $v_{\rm y}$. 
Note that $v_{\rm y}$ represents the evolution of turbulence 
because it is not observed in 2-dimensional plasmoid-chain. 
The panel shows that 
the velocity dispersion saturates around 0.02c, 
which means that 
only 0.01\% of magnetic field energy transferred into kinetic energy of turbulence
\footnote{
The ratio between turbulence and magnetic field energy density can be written as: 
\begin{equation}
  \frac{\epsilon_{\rm turb}}{\epsilon_{\rm B}} \simeq \frac{\alpha^2}{1 + \sigma}
  ,
\end{equation}
where $\epsilon_{\rm B} = B^2/8 \pi$, $\epsilon_{\rm turb} \sim \rho h v_{\rm turb}^2/2$, 
and $\alpha = v_{\rm turb}/c_{\rm A}$ 
when $v_{\rm turb} \ll c$.  
}
. 
On the other hand, 
the maximum velocity reaches around 30 to 40 \% of light velocity, 
indicating local appearance of very strong turbulence. 
Bottom panel of Figure \ref{fig:2} is the distribution of $v_{\rm y}$. 
We found that 
the distribution can be described by $f(|v_{\rm y}|) \propto \exp(-|v_{\rm y}|/\delta v_{\rm y})$ 
where $\delta v_{\rm y}$ is the velocity dispersion discussed above. 
Note that this is different from the usual turbulence 
whose 1-point velocity distribution function is usually the Gaussian. 

\begin{figure}
 \centering
  \includegraphics[width=8.cm,clip]{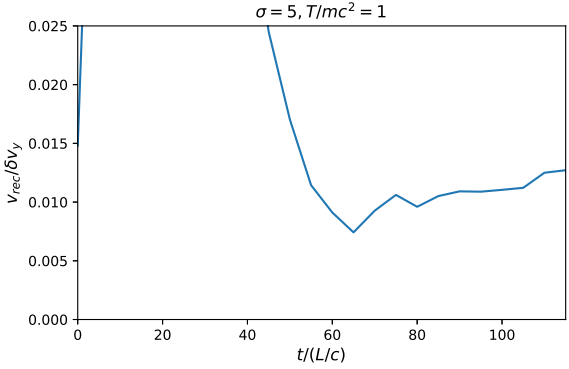}
  \caption{Temporal evolution of the ratio of reconnection rate $v_{\rm in}$ to the velocity dispersion of $v_{\rm y}$. 
           }
  \label{fig:2.2}
\end{figure}

Here we return to magnetic reconnection rate. 
\citet{1999ApJ...517..700L} found that 
if there is a sufficiently strong turbulence, 
magnetic reconnection rate can be described as:
\begin{equation}
  \frac{v_{\rm in}}{c_A} \simeq \mathrm{min}\left[ \left( \frac{L}{l} \right)^{1/2}, \left( \frac{l}{L} \right)^{1/2} \right] \left( \frac{v_l}{c_A} \right)^2 
  \label{eq:3.2}
  ,
\end{equation}
where $L$ is the sheet length, 
$l$ is the injection scale of turbulence, 
and $v_l$ is the injection velocity. 
This was proven to be valid in relativistic regime with a slight modification 
in the trans-Alfv\'enic regime \citep{2015ApJ...815...16T}. 
Unfortunately, 
direct application of the above equation is difficult in this case, 
because several parameters, $L, l, v_l$, and the numerical coefficient is difficult to know. 
Instead, 
the scaling of $v_{\rm in}$ in terms of $v_l$ would be useful for checking Equation (\ref{eq:3.2}). 
Figure \ref{fig:2.2} is a plot of the ratio of reconnection rate $v_{\rm in}$ to the velocity dispersion of $v_{\rm y}$. 
From $t = 60L/c$, 
the ratio becomes nearly constant after the turbulence started to evolve in the sheet. 
This indicates that 
Equation (\ref{eq:3.2}) needs a modification in this case 
if we reads $\delta v_{\rm y}$ as $v_l$. 
We consider that 
this is because the turbulence appeared in our simulation is different from the critical balanced one 
assumed for obtaining Equation (\ref{eq:3.2}). 
This can be seen in the distribution of $v_{\rm y}$ in Figure \ref{fig:2} 
which does not follow the usual Gaussian distribution, $\exp[-(v_{\rm y}/\delta v_{\rm y})^2]$, 
but $\exp[- |v_y|/\delta v_{\rm y}]$.
\footnote{
Although the temporal evolution of velocity dispersion in Figure \ref{fig:2} indicates that 
the turbulence reached a steady state, 
there is a possibility that 
the turbulence would evolve into critical-balanced turbulence if much longer time has passed. 
Our present numerical resources prohibits us to perform larger simulation, 
and it will be our future work to check the above statement. 
}
Note that 
the ratio shows a slight increase in that phase, 
and we consider that this would reflect an enhancement from local plasmoid-chain in the sheet. 

\section{Discussion}
\label{sec:sec5}

\subsection{Onset of 3-dimensional Instability}

\begin{figure}
 \centering
  \includegraphics[width=8.cm,clip]{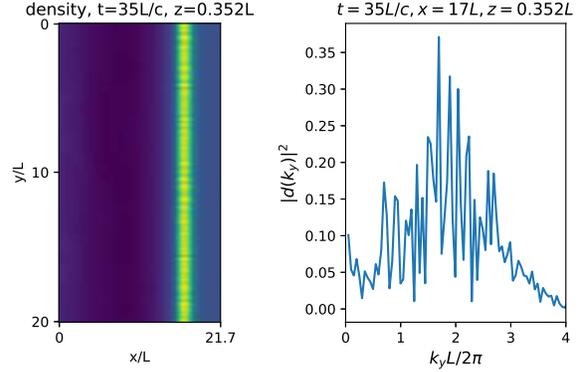}
  \caption{Density profile in the initial phase of the evolution of turbulence: $t = 35L$. 
           Left: density profile in a plane at $z = 0.352L$. 
           A yellow region is a plasmoid region. 
           Right: 1-dimensional profile of square of Fourier transformed density on a line at $x = 17L$ and $z = 0.352$ 
           which is along a plasmoid. 
           }
  \label{fig:3}
\end{figure}

In this section, 
the origin of turbulence in the sheet is discussed. 
The left panel of Figure \ref{fig:3} is a density profile in a plane located at $z = 0.352 L$ 
in the evolving phase of turbulence: $t = 35L/c$. 
The yellow region is a right-moving plasmoid. 
Note that 
a coherent structure in y-direction along the plasmoid can still be observed. 
However, 
a small density fluctuation in y-direction exists, 
indicating the onset of an instability. 
The right panel of Figure \ref{fig:3} is square of Fourier transformed density along y-direction 
at $x = 17L$ and $z = 0.352L$. 
It shows a peak at $k_{\rm y} L/2 \pi\sim 2$. 
This indicates that 
the density fluctuation does not result directly from the initial velocity perturbation with white noise, 
but some instabilities are responsible for it. 
In the following, 
we discuss the instability using the energy principle~\citep{2005ppfa.book.....K} 
which allows a simpler analysis than the usual linear analysis
\footnote{
In order to apply the energy principle analysis, 
the 0-th state should be a steady state. 
In our analysis, 
the background plasma is not a true steady state but transient state. 
However, 
the time scale of the background plasmoid evolution is much longer than that of the instability. 
Hence, we consider that 
it can be regarded as a quasi-steady state, 
and we apply the energy principle analysis on it. 
}. 
For simplicity, 
we assume that a plasmoid is static and at rest
\footnote{
We consider that 
this is not so bad approximation 
because the velocity of the plasmoid is sufficiently small comparing with the Alfv\'en velocity: $v_{\rm plasmoid} \ll c_{\rm A}$, 
and also the relativistic effects from the Lorentz factor is negligible. 
}
. 
The energy principle discuss 
the 2nd-order potential energy change by a displacement $\xi$, 
which is given as: 
\begin{align}
  \delta W^{(2)} &= \frac{1}{2} \int d x^3 \left[ 
    \frac{Q^2}{4 \pi} + {\bf J} \cdot (\vec{\xi} \times {\bf Q}) + \Gamma p_{\rm g} (\nabla \cdot \vec{\xi})^2 \right.
    \nonumber               
    \\
     &+ \left. (\vec{\xi} \cdot \nabla p_{\rm g}) (\nabla \cdot \vec{\xi})
    - \vec{\xi} \cdot \nabla \phi \nabla \cdot (e \vec{\xi})
  \right]
  ,  
  \label{eq:4.1}
\end{align}
where ${\bf Q} \equiv \nabla \times (\vec{\xi} \times {\bf B})$, 
${\bf J} = \nabla \times {\bf B}/4 \pi$ is the current density, 
and $e$ is the energy density. 
$\phi$ is a gravitational potential, 
which is assumed to be very small enough to allow us to apply non-relativistic treatment.  
The goal of the analysis is to find a displacement vector $\vec{\xi}$ 
which makes $\delta W^{(2)}$ negative. 
In this case, 
the system is unstable. 
On the other hand, 
the system is stable 
if $\delta W^{(2)}$ is positive for all the $\vec{\xi}$. 

\begin{figure}
 \centering
  \includegraphics[width=8.cm,clip]{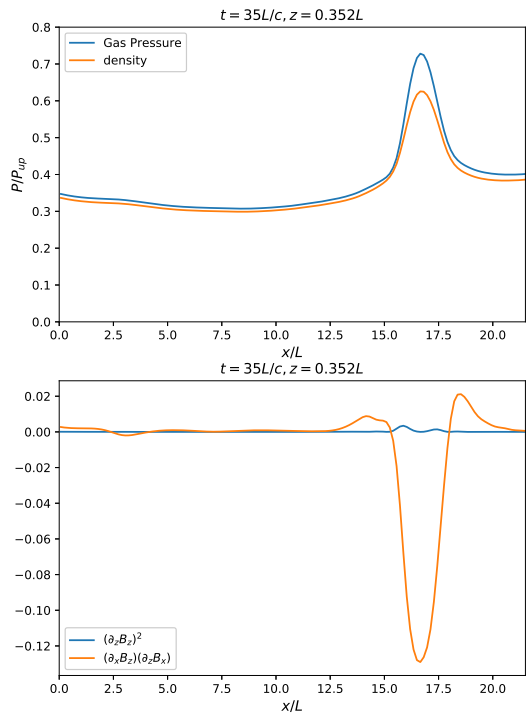}
  \caption{
           Profiles along 1-dimensional cut along $y = 10L, z = 0.352L$ at $t = 35L/c$, 
           which corresponds to a line passing upstream plasma and a part of plasmoid 
           in the early phase of their development. 
           Top: gas pressure and density. Bottom: spatial gradients of magnetic field. 
           }
  \label{fig:5}
\end{figure}

For simplicity, 
we consider the following trial displacement vector: 
\begin{align}
  &\nabla \cdot \vec{\xi} = 0
  , 
  \\
  &\vec{\xi} = (- \xi_{\rm x}(y), 0, 0)
  . 
\end{align}
To model the magnetic field structure around the plasmoid, 
we consider the following magnetic field: 
\begin{equation}
  {\bf B} = (B_x(x,z), 0, B_z(x,z))
  .
\end{equation}
Although we neglected the velocity of plasmoid, 
we take into account the acceleration of plasmoid as: $- \nabla \phi \rightarrow - \tilde{g} {\bf e}_{\rm x}$ 
where ${\bf e}_{\rm x}$ is the unit vector in x and $\tilde{g}$ is acceleration of the plasmoid. 
Then, 
Equation (\ref{eq:4.1}) reduces to
\begin{align}
  \delta W^{(2)} = \frac{1}{8 \pi} \int d x^3 \xi_{\rm x}^2 \left[ 
    (\partial_{\rm z} B_{\rm z})^2 + (\partial_{\rm x} B_{\rm z}) (\partial_{\rm z} B_{\rm x}) 
    - \tilde{g} (\partial_{\rm x} e)
  \right]
  . 
  \label{eq:4.2}
\end{align}
Figure \ref{fig:5} are the profile of density, pressure, and the gradients of magnetic field in Equation (\ref{eq:4.2}). 
It shows that 
the 2nd-order potential energy is negative in the left-hand side of plasmoid, 
indicating that 
the plasmoid is unstable for a displacement in negative x-direction. 
We consider that 
this is a kind of interchange instability or Rayleigh-Taylor type instability. 
This is because there is no counter force to prevent corrugation of plasmoid. 
Note that the plasmoid is surrounded by magnetic field with no guide field. 
It is also known that 
such a plasmoid can also be unstable for kink type instability. 
Hence, 
we can conclude that 
the initial plasmoid coherent in y-direction is unstable for several instabilities, 
and it leaves a density perturbation behind the current sheets, 
resulting in a seed of turbulent motion of secondary plasmoids. 

To obtain an expression of the growth rate of this instability, 
we assume that the plasmoid located at the origin 
with azimuthal magnetic field: 
\begin{equation}
  {\bf B} = B_{\theta}(r) {\bf e}_{\theta}, 
  \label{eq:4.3}
\end{equation}
where $r$ is the radius of the plasmoid, 
and $\theta$ is the azimuthal angle. 
Equation (\ref{eq:4.2}) reduces to
\begin{align}
  \delta W^{(2)} = - \frac{1}{8 \pi} \int d x^3 \xi_{\rm x}^2 \left[ 
    \frac{B_{\theta} \partial_r B_{\theta}}{r} 
    + \tilde{g} (\partial_{\rm x} e)
  \right]
  . 
  \label{eq:4.4}
\end{align}
To proceed further, 
we assume that the plasmoid has a constant current density 
whose magnetic field can be described as:
\begin{align}
  B_{\theta}(r) = 
  \begin{cases}
    B_0 (r/R_0) & (\text{if} \ r < R_0),
    \\
    B_0 & (\text{if} \ r > R_0),
  \end{cases}
  \label{eq:4.5}
\end{align}
where $R_0$ is the radius of the plasmoid. 
For simplicity, 
we assume the pressure balance, 
and also assume the internal energy $e$ is proportional to the gas pressure as: 
\begin{equation}
  \label{eq:4.6}
  e \simeq \tilde{e} \left[ 1 - \frac{B_0^2}{8 \pi P_{\rm tot}} \left( \frac{r}{R_0} \right)^2 \right] \quad ({\rm if} \ r < R_0), 
\end{equation}
where $P_{\rm tot}$ is the total pressure in the upstream region. 
For the displacement vector $\xi_{\rm x}$, 
we assume the sinusoidal expression as $\xi_{\rm x} = \xi_0 \cos k y$. 
To make the analysis easier, 
we assume the displacement vector work only inside of the plasmoid, 
that is, $\xi_{\rm x} = \xi_0 \cos k y {\rm Hv}(R_0 - r)$ where ${\rm Hv}(x)$ is the Heaviside step function. 
Then, Equation (\ref{eq:4.4}) becomes 
\begin{align}
  \delta W^{(2)} = - \frac{\xi_0^2 L_{\rm y} B_0^2}{16 \pi} (1 + \tilde{e} R_0 \tilde{g} )
  , 
  \label{eq:4.7}
\end{align}
where $L_{\rm y}$ is the size of sheet in y-direction. 
Note that here we performed the integral of x only in the negative-half plane ($x < 0$) 
to take into account the non-symmetric form of actual plasmoids 
which is equivalent to assuming the displacement vector only in the back-part of the plasmoid. 

Following \citet{2005ppfa.book.....K}, 
the approximate growth rate $\Gamma_{\rm grow}$ of the instability can be obtained as:
\begin{equation}
  \Gamma_{\rm grow} \sim \sqrt{- \frac{\delta W^{(2)}}{K}}
  ,
  \label{eq:4.8}  
\end{equation}
where
\begin{equation}
  K \equiv \frac{1}{2} \int d^3 x \rho h |\xi|^2
  .
  \label{eq:4.9}
\end{equation}
Here we modified the original non-relativistic expression of $K$ by multiplying the specific enthalpy $h$ 
because this term comes from the equation of motion. 
After a similar calculation, 
we obtain 
\begin{equation}
  \Gamma_{\rm grow} \sim \frac{c_A}{R_0} \sqrt{1 + \tilde{e} R_0 \tilde{g}}
  .
  \label{eq:4.10}  
\end{equation}
This means that 
this instability is driven by the Lorentz force and the effective acceleration, 
indicating the above discussion that interchange and Rayleigh-Taylor type instabilities are responsible. 
Note that the ratio of the Lorentz force and effective acceleration is absorbed in $\tilde{e}$. 
From dimensional analysis, 
Equation (\ref{eq:4.10}) also indicates that 
the characteristic wavelength in y-direction would be $R_0$, 
which reproduces the result in the right-panel of Figure \ref{fig:3}. 

This will also explain the small turbulent velocity shown in Figure \ref{fig:2}, 
because the source of the instability is the acceleration of plasmoid and the difference of energy in the upstream and inside of plasmoid. 
Both of them have small energy compared with the inertia of the hot plasma in plasmoid. 
Note that 
it is known that 
plasmoid is unstable for another several instabilities, 
such as the kink instability \citep{2000mrp..book.....B}. 
In our simulations, 
it is found that 
the above instability leaves strong perturbations in the current sheet, 
resulting in the seeds non-coherent secondary tearing instability and plasmoids in y-direction; 
Such a successive process finally generates the turbulent sheet. 
\footnote{
In the case of relativistic collisionless pair-plasma, 
the relativistic drift-kink instability can be a seed of 3-dimensional turbulence when the sheet thickness is close to the Debye length 
\citep{2005PhRvL..95i5001Z,2007ApJ...670..702Z,2014ApJ...783L..21S,2015PhRvL.114i5002L,2015ApJ...806..167G}. 
Although we consider the magnetohydrodynamic approximation, that is, much denser plasma and larger scale physics in this paper, 
those plasma scale fluctuations can also be a seed of perturbations in magnetohydrodynamic scale phenomena, 
which support our statement that current sheets in 3-dimensional space naturally evolve into turbulence. 
Note that such collisionless effects become important in low density plasma, such as pulsar magnetosphere and wind regions. 
}

\subsection{Difference Between 2 and 3 Dimensional Cases}

\begin{figure}
 \centering
  \includegraphics[width=8.cm,clip]{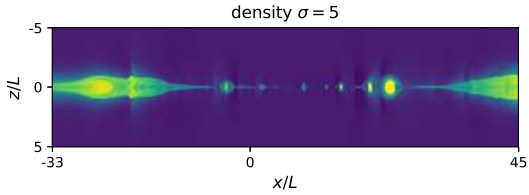}
  \caption{
           Density profile on x-z plane at $t = 120 L/c$ 
           obtained by a 2-dimensional simulation with the same set up as the 3-dimensional one. 
           }
  \label{fig:6}
\end{figure}

In 2-dimensional case, 
it is known that 
current sheets in high Lundquist number flow evolve into plasmoid-chain as discussed in Section \ref{sec:sec3}. 
Figure \ref{fig:6} is the temperature profile of plasmoid chain obtained in our 2-dimensional calculation. 
As is known, 
the sheet is filled with a lot of plasmoid, 
and a self-similar structure seems to be observed, 
allowing a locally very thin sheet. 
On the other hand, 
Figure \ref{fig:3} shows that 
plasmoid-chain does not appear in 3-dimensional case, 
and almost all the sheet is filled with self-generated turbulence. 

\begin{figure}
 \centering
  \includegraphics[width=8.cm,clip]{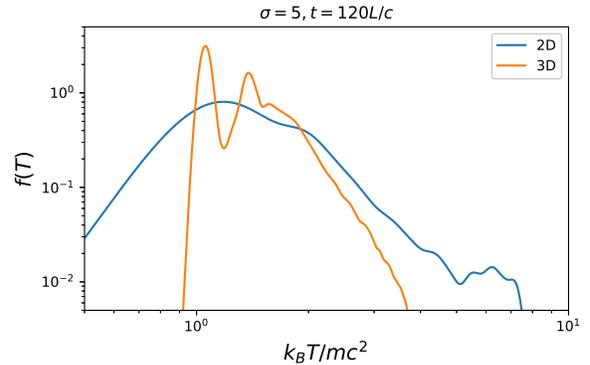}
  \caption{
           Distribution function of temperature on a plane ($z = 0$) at a saturation time ($t = 120 L/c$). 
           Both 2 and 3-dimensional results are plotted. 
           }
  \label{fig:7}
\end{figure}

Figure \ref{fig:7} is the temperature distribution in the plane located at the center of current sheet ($z=0$). 
It shows that 
3-dimensional turbulent sheet is much less active than 2-dimensional plasmoid-chain, 
and the temperature distribution becomes much narrower than the plasmoid-chain case. 
This is because 
plasmoid-chain results in very small plasmoid, 
and they experience collisions to neighboring plasmoids with approximately Alfv\'en velocity, 
resulting in frequent energy conversion from its kinetic velocity into thermal energy. 
On the other hand, 
3-dimensional turbulent sheets prohibits the appearance of such small plasmoids, 
and the energy concentrated in small plasmoids are distributed into turbulent motions. 
This indicates that 
relativistic current sheets are not a good candidate for small flares 
observed in many high energy astrophysical phenomena, 
because of no existence of small plasmoids and origin of small scale strong fluctuations. 
However, 
in our simulations 
the initially generated plasmoids can survive for a long time, 
and evolve into a very large ones, so called ``\textit{monster plasmoids}''~\cite{2010PhRvL.105w5002U,2013ApJ...775...50T,2014ApJ...783L..21S}. 
We consider that 
it is still possible to explain large flare phenomena by the appearance of the monster plasmoids. 


\section{Conclusion}
\label{sec:sec6}

In this paper, 
we studied an evolution of 3-dimensional relativistic current sheets in Poynting-dominated plasma. 
In 2-dimensional case, 
it is known that 
sufficiently long current sheets always evolve into plasmoid-chain, 
and their reconnection rate is around 0.01. 
The plasmoid-chain is found to be one of the final steady states in 2-dimensional cases, 
and observed in both of magnetohydrodynamic and collisionless simulations from non-relativistic to relativistic plasma. 
For this reason, 
the plasmoid-chain is well-studied and even applied for several astrophysical phenomena, such as the solar flares and flares in relativistic jets. 
However, 
the plasmoid-chain is a 2-dimensional phenomenon, 
and there is a doubt on the applicability in 3-dimensional cases. 
Actually, a few non-relativistic simulations reported that 
the plasmoid-chain was not observed but instead their sheets evolved into turbulence. 

In our simulation, 
we found that 
relativistic current sheets also develop into turbulence, 
and no clear evidence of plasmoid-chain was observed. 
The reconnection rate is around 0.004 
which is much smaller than that of plasmoid-chain, $0.03$. 
Using the energy principle, 
we also found that 
the plasmoid is unstable for an displacement of plasma in the opposite directions of plasmoid velocity, 
seemingly the interchange type instability, 
resulting in turbulence in the sheet. 
Finally, 
we discussed the activity of the sheet using the temperature distribution in it. 
It is found that 
the 3-dimensional sheet has the smaller number of high temperature regions than the 2-dimensional sheet. 
This is because the small scale plasmoids responsible for generating high temperature region do not appear in turbulent sheet, 
and the turbulence itself also increase the sheet width by its eddy motion.  
This indicates that 
the realistic relativistic current sheet in Poynting-dominated plasma is not so active, 
and cannot be a candidate for mini-flares observed in many high-energy astrophysical phenomena. 
However, 
the limitation of our numerical resources and high numerical costs of 3-dimensional simulations 
prohibits us from the sufficient parameter search, in particular, the dependence on the $\sigma$-parameter, 
and it still remains unknown if the sheet with higher $\sigma$-parameter upstream plasma would be more active. 
And this is our future work. 

\section*{Acknowledgments}
We would like to thank Masahiro Hoshino, Takanobu Amano, Masanori Iwamoto, Shuichiro Inutsuka, and Tsuyoshi Inoue 
for many fruitful comments and discussions. 
Numerical computations were carried out on the Cray XC30 
at Center for Computational Astrophysics, CfCA, of National Astronomical Observatory of Japan.
This work is supported by the Postdoctoral Fellowships by the Japan Society for the Promotion of Science No. 201506571 (M. T.). 

\appendix

\section{Resolution Study}

In the main part of this paper, 
we performed a numerical simulation of relativistic current sheets. 
This simulation demands us a very high resolution 
because of the development of turbulence in current sheet scale. 
In this Appendix, 
we discuss the effect of resolution. 

\begin{figure}
 \centering
  \includegraphics[width=8.cm,clip]{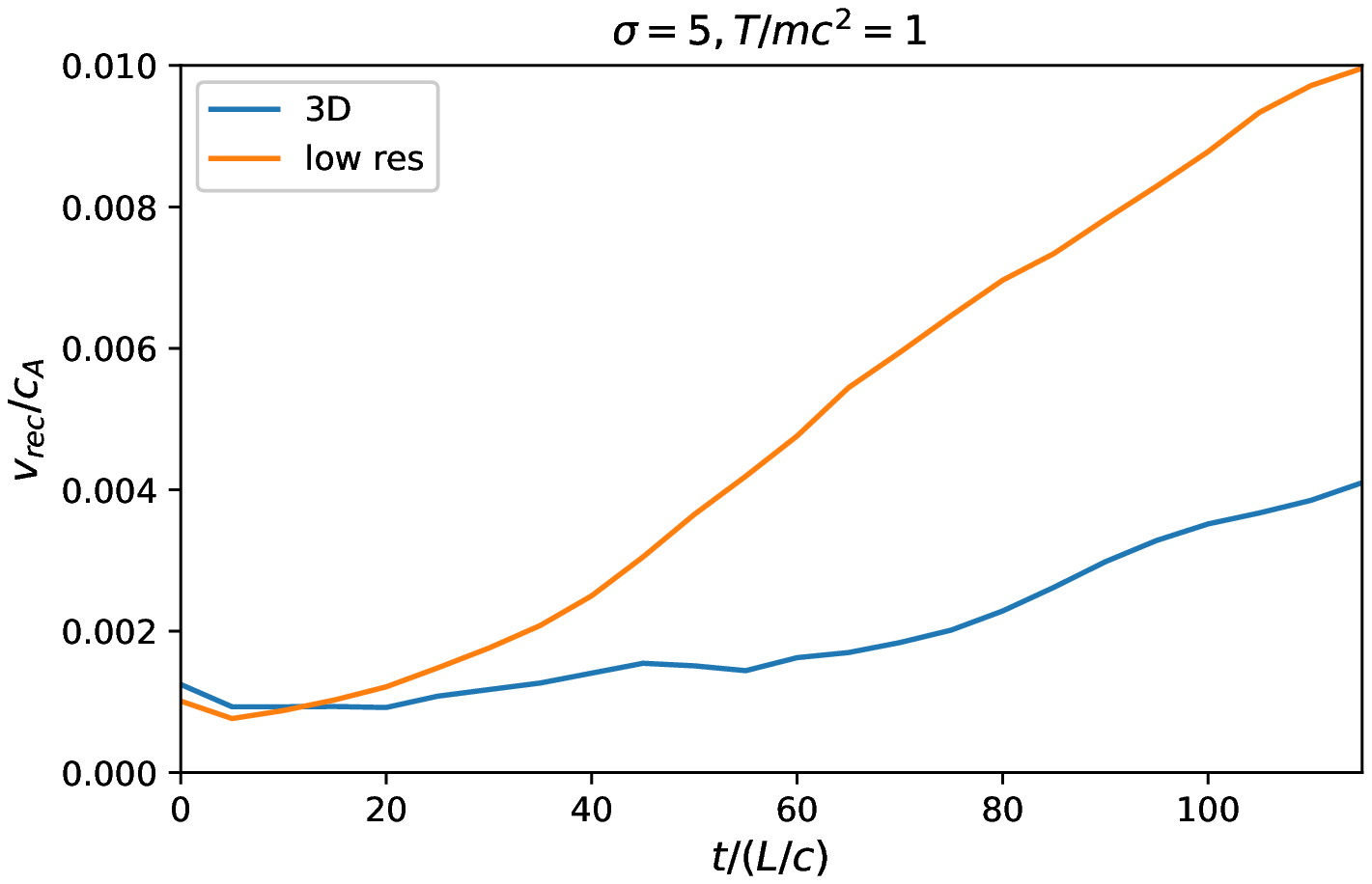}
  \includegraphics[width=8.cm,clip]{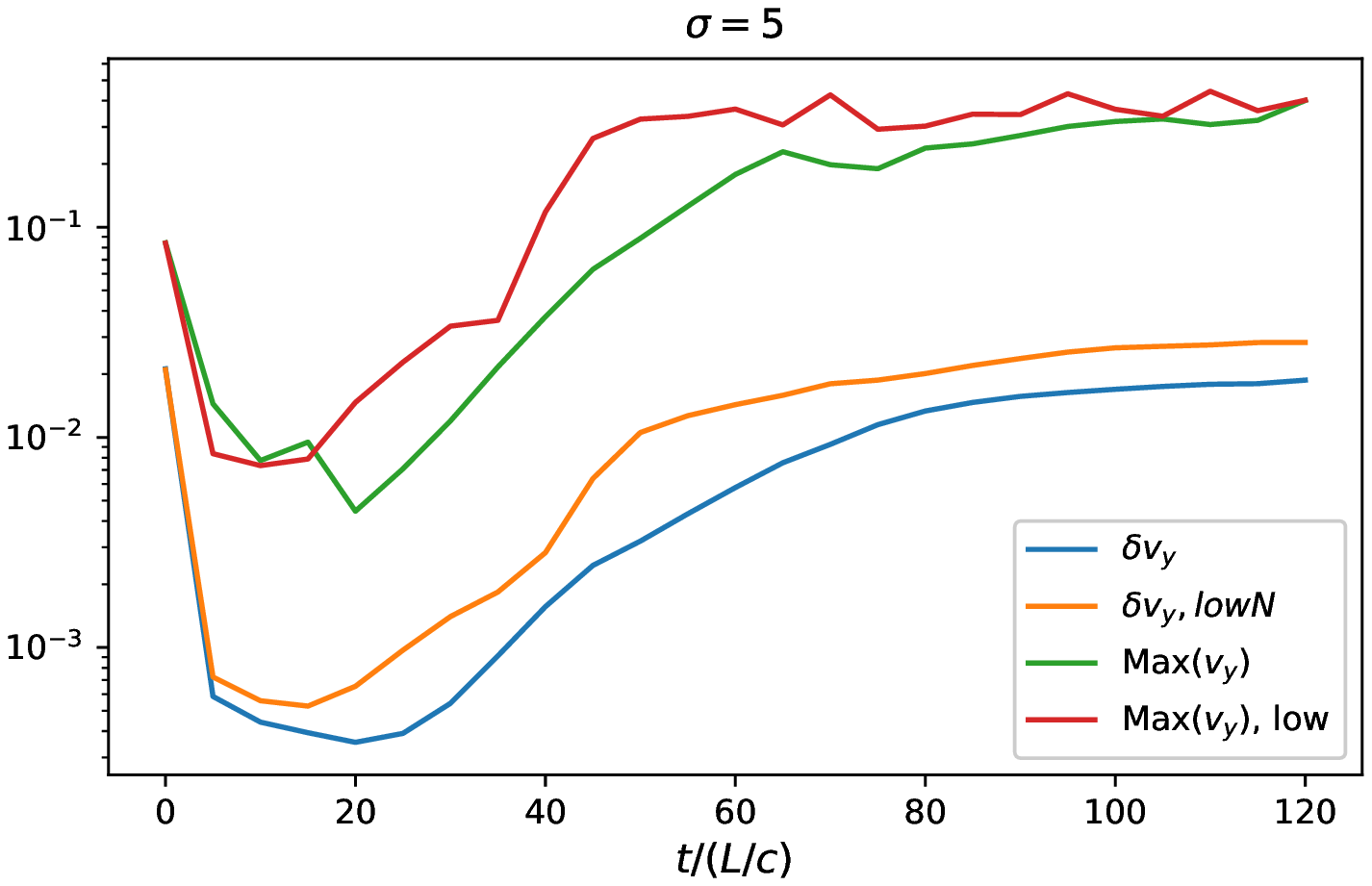}
  \caption{A numerical results of a simulation with the same setup but used half mesh number in all directions. 
           Top: reconnection rate, 
           Bottom: maximum and dispersion of $v_{\rm y}$. 
           }
  \label{fig:A1}
\end{figure}

Figure \ref{fig:A1} is a numerical results of the same setup but used half mesh number in all directions, 
that is, $N_{\rm x} \times N_{\rm y} \times N_{\rm z} = 2048 \times 512 \times 1024$. 
The top panel of the figure is the reconnection rate. 
It shows that 
the lower resolution run shows much larger reconnection rate than the higher resolution run. 
The bottom panel of the figure is the maximum and dispersion of $v_{\rm y}$. 
It indicates that 
the saturation level of maximum and dispersion of $v_{\rm y}$ is similar, 
though the onset of the growth is a little faster in the lower resolution run. 
Note that this is consistent with \citep{2017ApJ...838...91K}. 
We consider that 
the higher reconnection rate in the lower resolution run does not result from physical effects but numerical effects. 
This is because the turbulence appeared in the simulation is very similar, 
indicating that reconnection rate is not controlled by the turbulence in the sheet. 
Note that 
the resistivity in the lower resolution simulation is only a little less than the numerical resistivity, 
and this also allowed the strong numerical effects on local magnetic reconnection in a region 
where the sheet width becomes close to the numerical cell size. 
Unfortunately, 
the present numerical resource does not allow us to check a higher resolution study. 

\section{Numerical Scheme}

In this section, 
our new method for RRMHD is explained. 
For simplicity, 
only the 1-dimensional case is discussed for the treatment of numerical flux. 
First, the basic equations are discussed. 
In the RRMHD case, 
the Maxwell equations are considered: 
\begin{align}
\nabla \cdot \mathbf{E} &= q
\label{eq:Maxwell_1}
,
\\
\nabla \cdot \mathbf{B} &= 0
\label{eq:Maxwell_2}
,
\\
\partial_t \mathbf{E} - \nabla \times \mathbf{B} &= - \mathbf{J}
\label{eq:Maxwell_3}
,
\\
\partial_t \mathbf{B} + \nabla \times \mathbf{E} &= \mathbf{0}
\label{eq:Maxwell_4}
,
\end{align}
where ${\bf B} \equiv {\bf B}/\sqrt{4 \pi}$ and ${\bf E} \equiv {\bf E}/\sqrt{4 \pi}$ in the Gauss unit. 
In addition, 
the plasma part can be described by: 
\begin{align}
\partial_t
\left(
 \begin{array}{c}
   D \\
   m^i \\
   e
 \end{array}
\right)
+ \partial_j
\left(
 \begin{array}{c}
   F_D^j \\
   F_m^{ij} \\
   F_e^j
 \end{array}
\right)
= 0,
\label{eq:fluid}
\end{align}
where the conserved variables are: 
\begin{align}
D &= \gamma \rho
\label{eq:D}
, \\
\mathbf{m} &= \rho h \gamma^2 {\bf v} + \mathbf{E \times B}
\label{eq:m}
, \\
e &= \rho h \gamma^2 - p + \frac{1}{2} (E^2 + B^2)
\label{eq:e}
,
\end{align}
where ${\bf v}$ is the fluid three-velocity, $\gamma = (1 - v^2)^{-1/2}$ is the Lorentz factor, 
and numerical fluxes are: 
\begin{align}
F_D^i &= D v^i 
,
\\
F_m^{ij} &= m^i v^j + p \eta^{ij}
- E^i E^j - B^i B^j 
+ \frac{1}{2} (E^2 + B^2) \eta^{ij}
,
\\
F_e^i &= m^i
,
\end{align}
where $\eta^{\mu \nu}$ is the metric tensor. 
Although obtaining charge density $q$ and current density ${\bf J}$ is non-trivial, 
which has already been explained in our previous work \citep{2011ApJ...735..113T}. 
In the following, 
we assume the charge density and current density are determined by some method, 
and concentrate on obtaining numerical flux. 
In the above equations, 
we have to update the following variables: $\{D, m_i, e, {\bf E}, {\bf B} \}$. 
To obtain the Godunov-type exact solution, 
we have to solve the jump condition on the above equation. 
However, it is too complicated to solve, 
and there is no study succeeding in obtaining exact solutions of RRMHD full Godunov scheme. 
Here, 
we propose a better method allowing us to reproduce the exact RMHD numerical flux in ideal regime 
and take into account resistivity, 
though it is not the exact numerical flux of RRMHD equation itself. 
In our new scheme, 
the numerical flux ${\bf F}$ is calculated using HLLEM Riemann solver \citep{2016JCoPh.304..275D} as: 
\begin{align}
  \label{eq:B1}
  {\bf F}_{\rm HLLEM} &\equiv \frac{s_{\rm R} {\bf F}_{\rm L} - s_{\rm L} {\bf F}_{\rm R}}{s_{\rm R} - s_{\rm L}} 
         + \frac{s_{\rm R} s_{\rm L}}{s_{\rm R} - s_{\rm L}}  ( {\bf Q}_{\rm R} - {\bf Q}_{\rm L})
  \nonumber
  \\
  &- \varphi \frac{s_{\rm R} s_{\rm L}}{s_{\rm R} - s_{\rm L}} {\bf R}_*(\bar{\bf Q}) {\bf \delta}_*(\bar{\bf Q}) {\bf L}_*(\bar{\bf Q})( {\bf Q}_{\rm R} - {\bf Q}_{\rm L})
  ,
\end{align}
where ${\bf Q}$ is the conservative variables, and ${\rm F}$ is the flux. 
The subscript L and R mean the left-hand and right-hand side variables. 
$s_{\rm L/R}$ is the fastest characteristic speed in left and right directions. 
The first line is the classical HLL flux; 
the second line is the ``anti-diffusion term'' which subtracts numerical dissipation stemming from characteristic mode neglected in HLL flux. 
The detailed explanation is given in the original paper \citep{2016JCoPh.304..275D}. 
In our scheme, 
we calculate the anti-diffusion term using the RMHD eigen-values 
which has already been obtained, for example, by \citet{2010ApJS..188....1A}. 
Hence, for the electric field $\{ {\bf E} \}$, 
the numerical flux is calculated by classical HLL flux, 
and for the other variables $\{D, m_i, e, {\bf B} \}$, 
the numerical flux is calculated by the HLLEM flux. 
This allows us to include the internal structure of RMHD in the intermediate HLL state. 
Note that the resistivity is naturally taken into account from the non-ideal RMHD terms in the above equations (\ref{eq:D})-(\ref{eq:e}). 

\begin{figure}
 \centering
  \includegraphics[width=8.cm,clip]{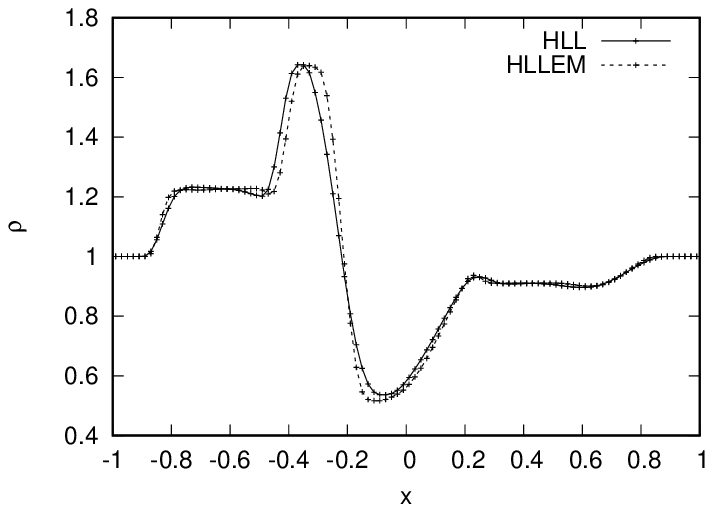}
  \includegraphics[width=8.cm,clip]{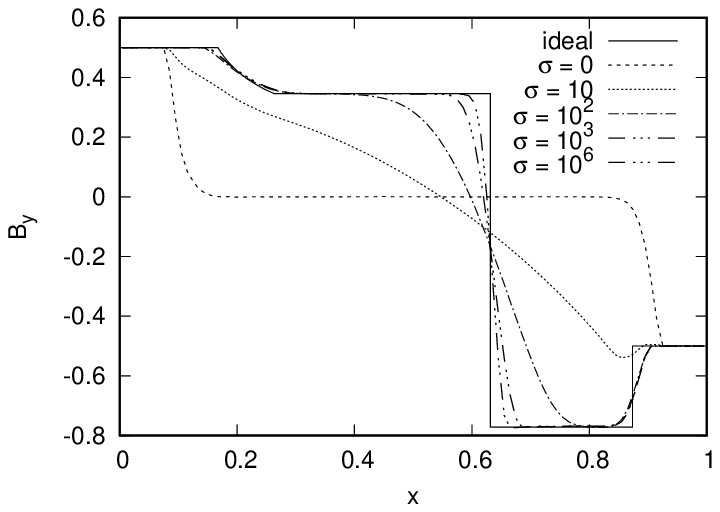}
  \caption{A numerical results of shock tube problems. 
           Top: Scheme dependence. 
           Bottom: electric conductivity dependence. 
           }
  \label{fig:B1}
\end{figure}

In the following, 
some numerical results obtained by our new scheme will be shown, 
and discuss the accuracy and validity of the scheme. 
$\Gamma = 4/3$ is assumed for all the problems, 
and grid number $N=100$ is used. 
Spatial and temporal 2nd-order accuracy scheme is assumed. 
Figure \ref{fig:B1} is the numerical results of shock tube problems. 
In the left panel, 
a problem including slow shocks is considered. 
The initial left and right states are given as: 
\begin{align}
(\rho^L, (v_z)^L, p^L, (B^y)^L) =& (1, 0.2, 10^{-2}, 0.2)
\\
(\rho^R, (v_z)^R, p^R, (B^y)^R) =& (1, -0.2, 5 \times 10^{-2}, -0.1)
\end{align}
All the other fields are set to $0$. 
The electric conductivity $\sigma \equiv c^2 / \eta$ is set to $10^6$ 
to reproduce ideal results. 
The left panel of Figure \ref{fig:B1} is the density profile of our numerical results of HLL and HLLEM scheme at $t = 3.2$. 
It shows that 
the new HLLEM scheme allows us to capture the contact discontinuity ($x=-0.2$) and slow shock ($x=-0.5$) comparing with HLL scheme. 
Note that the new scheme also allows us to capture the fast shock ($x=-0.8$) by smaller number of grid points. 

The right panel of Figure \ref{fig:B1} is the simple MHD version of the Brio and Wu test. 
The initial left and right states are given by 
\begin{align}
(\rho^L, p^L, (B^y)^L) =& (1.0, 1.0, 0.5) 
\\
(\rho^R, p^R, (B^y)^R) =& (0.125, 0.1, -0.5)
\end{align}
All the other fields are set to $0$. 
The panel shows the numerical results of the same problem 
that changes the conductivity $\sigma = 0, 10, 10^2, 10^3, 10^6$. 
We also plot the exact ideal RMHD solution by the solid line 
computed by a publicly available code developed by Giacomazzo and Rezzolla~\citep{2006JFM...562..223G}. 
This result shows that 
our numerical solution reproduces the results in \citep{2011ApJ...735..113T}, 
and the new method can take into account resistive effect properly. 


\bsp

\label{lastpage}

\end{document}